\documentclass[a4paper]{article}
\usepackage[intlimits]{amsmath}
\usepackage{amsmath}
\usepackage{amsfonts}
\usepackage{amssymb}
\usepackage[dvips]{graphicx}

\begin{document}

\title{\textbf{Statistics of conductance and shot-noise power for chaotic cavities}}
\author{H.-J. Sommers$^1$, W. Wieczorek$^1$ and D. V. Savin$^2$}
\maketitle
\begin{center}
$^1$Fachbereich Physik, Universit\"{a}t Duisburg-Essen,
47048 Duisburg, Germany \\
$^2$Department of Mathematical Sciences, Brunel University, \\ Uxbridge, UB8
3PH, UK
\end{center}
\begin{abstract}
We report on an analytical study of the statistics of conductance,
$g$, and shot-noise power, $p$, for a chaotic cavity with arbitrary
numbers $N_{1,2}$ of channels in two leads and symmetry  parameter
$\beta = 1,2,4$. With the theory of Selberg's integral the first
four cumulants of $g$ and first two cumulants of $p$ are calculated
explicitly. We give analytical expressions for the conductance and
shot-noise distributions and determine their exact asymptotics near
the edges up to  linear order in distances from the edges. For
$0<g<1$ a power law for the conductance distribution is exact. All
results are also consistent with numerical simulations.
\\[2ex]
\hspace*{5ex}PACS numbers: 73.23.-b, 73.50.Td, 05.45.Mt, 73.63.Kv
\end{abstract}

\section{Introduction}

Our goal is to discuss the statistics of conductance and shot-noise
power for chaotic cavities using essentially the properties of
Selberg's integral. The static conductance $G$ relates linearly the
time averaged current $\overline{I(t_0)}= G\cdot  V$ to the external
voltage $V$ between two electron reservoirs. Fluctuations of the
current around its mean value are conventionally described by the
spectral noise power $P = 2 \int_0^{\infty} dt \overline{\delta
I(t+t_0) \delta I(t_0)}$, with $\delta I = I-\overline{I}$. As
temperature goes to zero, the only source of noise that remains
non-vanishing is related to the discreteness of the electric charge
carriers, the so-called shot-noise. For mesoscopic conductors, an
adequate framework for the problematic is based on
random-matrix-theory (RMT) approach to quantum transport, we refer
to \cite{Beenakker,Blanter} for reviews.

We consider a chaotic cavity with two attached leads supporting $N_1$ and
$N_2$ channels, respectively, and coupled perfectly to the interior of the
cavity. According
to the Landauer-B\"{u}ttiker formalism, 
the conductance $G$ and the shot-noise power $P$ can be expressed in
terms of the transmission eigenvalues $T_i$ as follows:
\begin{equation}\label{g}
 G = G_0 \sum_{i=1}^{n} T_i \equiv G_0 \cdot g
\end{equation}
and
\begin{equation}\label{p}
 P = P_0 \sum_{i=1}^{n} T_i (1-T_i) \equiv P_0 \cdot p\,,
\end{equation}
where $G_0 = \frac{2 e^2}{h}$ is the conductance quantum and $P_0 =
2 e V G_0$. The positive numbers $T_i\leq1$ are the
$n\equiv\mathrm{min}(N_1,N_2)$ non-zero eigenvalues of the matrix
$tt^{\dagger}$, with the transmission matrix $t$ being the
$N_1\times N_2$ submatrix of the full scattering matrix and
consisting of transition amplitudes from $N_1$ left to $N_2$ right
channels. For chaotic cavities, universal fluctuations of $T_i$ can
be described by  RMT \cite{Beenakker}.

Considering the moments of $g$ and $p$, the exact (RMT) results valid at
arbitrary channel numbers $N_{1,2}$ and repulsion parameter $\beta$ were
reported in the literature only for the average and variance of the
conductance \cite{Baranger,Jalabert,Beenakker}:
\begin{equation}\label{<g>}
\langle g \rangle = \frac{N_1 N_2}{N -1 + \frac{2}{\beta}}\,, \qquad N\equiv
N_1+N_2\,,
\end{equation}
\begin{equation}\label{varg}
\mbox{var}(g) = \frac{2}{\beta}
\frac{N_1N_2(N_1-1+\frac{2}{\beta})(N_2-1+\frac{2}{\beta}) }{
(N-2+\frac{2}{\beta})(N-1+\frac{2}{\beta})^2 (N-1+\frac{4}{\beta})} \,,
\end{equation}
and very recently for the average shot-noise power \cite{Savin}:
\begin{equation}\label{<p>}
\langle p \rangle =
\frac{N_1N_2(N_1-1+\frac{2}{\beta})(N_2-1+\frac{2}{\beta}) }{
(N-2+\frac{2}{\beta})(N-1+\frac{2}{\beta})(N-1+\frac{4}{\beta})} =
N_1N_2\frac{\beta}{2}\frac{\mathrm{var}(g)}{\langle g \rangle}\,.
\end{equation}
To derive these results in a uniform way, it is convenient  to use
 the known expression for the joint probability density
of transmission eigenvalues $T_i$ \cite{Savin,Beenakker}
\begin{equation}\label{jpd}
\mathcal{P}(T_1,T_2,\ldots,T_n) = \mathcal{N}_\beta^{-1} \prod_{i=1}^n
T_i^{\alpha -1} \prod_{j<k}|T_j - T_k |^\beta\,,\quad
\alpha\equiv\frac{\beta}{2}(|N_1-N_2|+1)\,,
\end{equation}
to perform the corresponding integrations on Eqs. (\ref{g})--(\ref{p}). The
normalization constant $\mathcal{N}_\beta$ above is given by
\begin{equation}
\mathcal{N}_\beta = \prod_{j =1}^{n-1} \frac{\Gamma (1+ \frac{\beta}{2} (1+
j)) \; \Gamma (\alpha + \frac{\beta}{2} j)\; \Gamma (1 + \frac{\beta}{2} j)
}{\Gamma (1 + \frac{\beta}{2})\; \Gamma ( 1+ \alpha + \frac{\beta}{2} (n + j
- 1))}
\end{equation}
and assures that (\ref{jpd}) is a probability density. It is known
for discrete positive $n$ and continuous $\alpha$ and $\beta$ as
Selberg's integral \cite{Mehta}.

As to the distribution functions, simple closed expressions can be obtained
for the conductance distribution at $n=1,2$
\cite{Baranger,Jalabert,Garcia-Martin} and for the shot-noise distribution at
$N_1=N_2=1$ only \cite{Pedersen}. To the best of our knowledge, no general
results valid at arbitrary $N_{1,2}$ and $\beta$ have been presented thus
far.

\section{Cumulants}

To study the cumulants of $g$ and $p$, one needs to know what are the
moments $\langle T_1^{n_1} \cdots T_k^{n_k}\rangle$, with
$\langle\ldots\rangle$ standing for the integration over the joint
probability density (\ref{jpd}) and $n_i\geqslant0$. Moments with all $n_i =
1$ as well as $\langle T_1^2\rangle$ can be found from recursion relations
already given in Mehta's book \cite{Mehta} that come from the Selberg's
integral theory. The latter can be successfully applied \cite{Savin} to
derive results (\ref{<g>})--(\ref{<p>}). This approach was recently extended
further to find $\langle T_1^3\rangle$ and $\langle
T_1(1-T_1)T_2(1-T_2)\rangle$ exactly \cite{Novaes2007}. Here, we have
calculated all moments with $\sum_i n_i \leqslant 4$ by using tricks of
partial integrations to reduce all moments to forms of Selberg's integral.
We will report on that in more detail elsewhere
\cite{Savin_Sommers_Wieczorek}. By this method we are able to calculate the
so-called skewness (third cumulant) of the conductance, which we represent
in the following compact form:
\begin{eqnarray}
\nonumber
 \langle \langle g^3 \rangle \rangle &\equiv& \langle (g-\langle g \rangle)^3 \rangle \\
 &=& \mathrm{var}(g) \frac{4[(1-\frac{2}{\beta})^2-(N_1-N_2)^2] }{
\beta(N-3+ \frac{2}{\beta}) (N-1+\frac{2}{\beta})(N-1+\frac{6}{\beta})}\,.
\end{eqnarray}
It is worth noting that the skewness vanishes for symmetric cavities
($N_1=N_2$) at $\beta=2$. This also holds for any odd cumulant of
the conductance, as the conductance distribution becomes symmetric
around $\frac{n}{2}$ in this case.\footnote{This can be seen from
the symmetry of the integral kernel (\ref{jpd}) at $\alpha=1$ by the
change of all $T_{i}\to1-T_i$ in $P_g(g)=\langle\delta(g-\sum_i
T_i)\rangle$.}

We have also calculated the kurtosis $\langle\langle g^4 \rangle\rangle$ of
the conductance (which is the fourth cumulant of $g$) and the variance
$\langle\langle p^2 \rangle\rangle=\mathrm{var}(p)$ of the shot-noise power
(the second cumulant). These expressions are given explicitly but are too
lengthy to be reported here. In the case of the single-mode leads,
$N_1=N_2=1$, one gets $\langle\langle g^4
\rangle\rangle=-\frac{32}{4725},-\frac{1}{120},-\frac{1}{540}$ and
$\mathrm{var}(p)=\frac{4}{525},\frac{1}{180},\frac{1}{180}$ at the values of
$\beta=1$, 2, 4, respectively. In the opposite semiclassical limit of large
channel numbers, $N_{1,2}\gg1$, we write the results as the following $1/N$
expansions:
\begin{eqnarray}
\nonumber \frac{\langle\langle g^4\rangle\rangle}{\mathrm{var}(g)} &=&
\frac{24 }{\beta^2 N^6}\Bigl[(N_1-N_2)^2 (N_1^2 + N_2^2 - 4 N_1 N_2) \\
\nonumber && + \frac{\beta-2}{\beta N} \Bigl( 12(N_1^4 + N_2^4) -
64N_1N_2(N_1^2 + N_2^2) + 105 N_1^2 N_2^2 \Bigr) \Bigr] \\
&& + {\cal O} (1/N^4)\,, \label{<g^4>}
\\ \nonumber
\frac{\mathrm{var}(p)}{\langle p \rangle} &=&
\frac{2}{\beta N^5} \Bigl[ N_1^4 + N_2^4 - 4 N_1 N_2 (N_1-N_2)^2 \\
\nonumber && + \frac{\beta-2}{\beta N} \Bigl( 9(N_1^4 + N_2^4) -
42N_1N_2(N_1^2 +  N_2^2) + 70N_1^2 N_1^2 \Bigr) \Bigr] \\
&& + {\cal O} (1/N^3)\,. \label{varp}
\end{eqnarray}
One can readily see that higher cumulants contribute at least in the next
order of $\frac{1}{N}$, so that the full distributions get more
Gaussian-like as $N$ grows. This tendency becomes even stronger for the
conductance distribution in symmetric cavities at $\beta=2$, as
$\langle\langle g^4\rangle\rangle$ vanishes then in the leading and
next-to-leading orders. In this case of symmetric cavities, $N_{1,2}=n\gg1$,
we can further find:
\begin{eqnarray}
\langle\langle g^4\rangle\rangle &=& \frac{3}{128\beta^3n^3}
\biggl(1-\frac{2}{\beta} + \frac{(\beta+2)^2}{2\beta^2n}\biggr)
+\mathcal{O}\Bigl(\frac{1}{n^5}\Bigr)\,,
\\
\mathrm{var}(p) &=& \frac{1}{64\beta} \biggl(1 + \frac{\beta-2}{\beta n}
+\frac{4+\beta(\beta-2)}{2\beta^2n^2}\biggr)
+\mathcal{O}\Bigl(\frac{1}{n^3}\Bigr)\,.
\end{eqnarray}

\section{Distributions}

Finally, we discuss shortly the distribution of the conductance,
$P_g(g)=\langle\delta(g-\sum_i T_i)\rangle$ and that of the
shot-noise power, $P_p(p)=\langle\delta(p-\sum_i T_i(1-T_i)\rangle$,
deferring a detailed consideration to a separate publication
\cite{Wieczorek_Savin_Sommers}. Writing the functions $P_g(g)$ and
$P_p(p)$ as Fourier series over the interval of support, we obtain,
e.g., for the conductance distribution the following representation:
\begin{equation}
\label{prob}
 P_g^{(\beta)} (g)  = n! \sum_{m=1}^\infty \frac{2}{n} \sin
\bigl( \frac{m \pi g}{n} \bigl) A^{(\beta)} (m)
\end{equation}
where $A^{(\beta)}(m)$ is known at $\beta = 1,2,4$. For example,
\begin{equation}
A^{(2)} (m) = C_2 \mbox{\,Im\,}\det B^{(2)}_{kl}(m)
\end{equation}
is given by the imaginary part of the determinant of the following matrix:
\begin{equation}
B^{(2)}_{kl}(m) = \int_0^1 dT \; T^{\alpha + k + l -3} {\rm e}^{i m \pi
T/n},\qquad \mbox{for}\quad k,l = 1,2, \ldots ,n \; .
\end{equation}
In a similar way we may write $P_g^{(1)}(g)$ and $P_g^{(4)}(g)$ as imaginary
parts of certain Pfaffians. The sum (\ref{prob}) can be done numerically
where care has to be taken for the Pfaffians, which may equivalently be
written as quaternion determinants of certain self-dual antisymmetric
matrices, which are easier to be calculated.

Analogous expressions are also obtained for $P_p^{(\beta)}(p)$.

\begin{figure}[b]
\centering
\includegraphics[width=7cm]{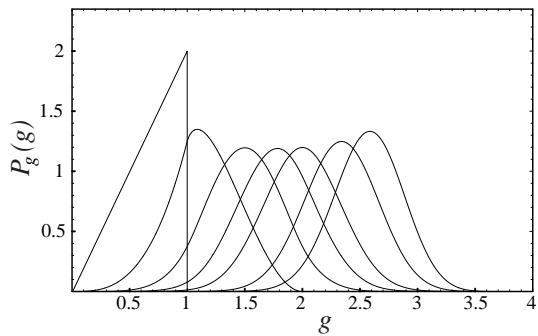}
\caption{\label{fig1} The conductance distribution at $\beta=1$, $N_1=4$ and
$N_2=1$, 2, 3, 4, 5, 7, 9 (from left to right).}
\end{figure}

It is worth noting that at any finite $n$ these distributions are continuous
but not analytic functions everywhere. Nonanalyticity in the distribution of
conductances in quasi-one-dimensional wires (the cusp point at $g=1$) was
recently reported in \cite{Garcia-Martin,Muttalib}, see also
\cite{Baranger}. In the present context of chaotic cavities, it can be
understood from the following geometrical consideration. In the case of
$P_g(g)$, calculating the average amounts to an integration over an
($n-1$)--di\-men\-sio\-nal plane $ g = \sum_i T_i$ cut by the
$n$--di\-men\-sional cube $0 \leqslant T_i \leqslant 1$, such that there
appear (sometimes weak) singularities at all points $g = M$ with an integer
$M$, $0\leqslant M \leqslant n$. A similar situation appears also for
$P_p(p)$. Here, we integrate over the ($n-1$)--di\-men\-sional sphere
$\frac{n}{4}-p=\sum_i(T_i-\frac{1}{2})^2$ and singularities appear whenever
the sphere touches one edge or surface of the cube, that is for
$p=\frac{M}{4}$ with $0\leqslant M \leqslant n$.

\begin{figure}[t]
\begin{center}
\includegraphics[width=12cm]{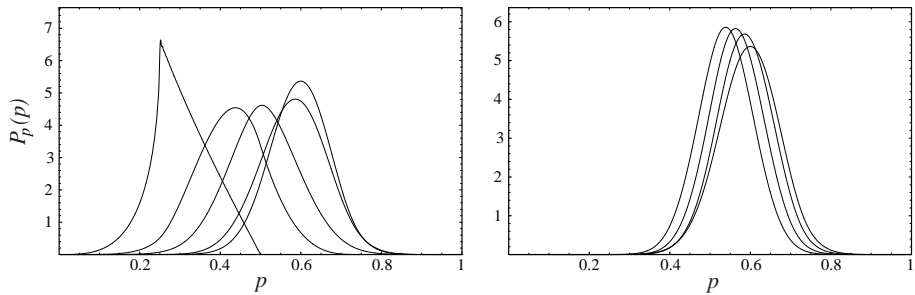}
\caption{\label{fig2} Distributions of the shot-noise power at $\beta=2$,
$N_1=4$ and $N_2=2$, 3, 4, 6, 8 (left plot) and at $\beta=2$, $N_1=4$ and
$N_2=8$, 10, 12, 14 (right plot).}
\end{center}
\end{figure}

Figure \ref{fig1} illustrates the conductance distribution at $\beta=1$
keeping $N_1=4$ fixed and varying $N_2$ from 1 to 9. In figure \ref{fig2}, we
plot the distribution of the shot-noise power at $\beta = 2$, keeping $N_1 =
4$ fixed and varying $N_2$ from 2 to 14. In both cases one can see the
tendency of the distributions to get peaked around the mean values, so that
in the bulk they can be effectively described by a Gaussian with the known
mean and variance, see Eqs. (\ref{<g>}) -- (\ref{<p>}) and (\ref{varp}).

\section{Asymptotics}

We are able to give some exact asymptotics near the edges of support of the
distributions expressed in terms of exact integrals, which are related to
some general forms of Selberg's integral. For example, for $0<g<1$ we find
\begin{equation}
P_g^{(\beta)}(g) = \mathrm{const}\times g^{\alpha
n+\frac{\beta}{2}n(n-1)-1}\,,
\end{equation}
where the proportionality factor is exactly known. This result follows
easily from calculating the powers of $g$ in $\langle\delta(g-\sum_i
T_i)\rangle$ under scaling of all $T_i\to g\widetilde{T}_i$ and noticing
that the upper integration limit of $\widetilde{T}_i$ remains unchanged
($=1$) as long as $g<1$. Similarly, $P_g^{(\beta)}(g)$ behaves near the
upper edge as follows:
\begin{equation}
P_g^{(\beta)}(g) \propto  (n-g)^{(n-1) (1+\frac{\beta}{2}n)}\,, \qquad n-1 <
g < n \,.
\end{equation}
For the distribution of the shot-noise power, one finds
\begin{equation}
P_p^{(\beta)}(p) \propto \left(\frac{n}{4} - p\right)^{\frac{n}{2} (1+
\frac{\beta}{2} (n-1)) -1} \,, \qquad \frac{n-1}{4} < p < \frac{n}{4} \,,
\end{equation}
whereas near the lower edge ($0<p<\frac{1}{4}$) the asymptotics is a bit
more complicated, since the contribution of the corners of the cube of
integration become disconnected. These expressions can be extended outside
the regions specified above, being valid then in a linear order in distances
from the edges. The expansions can even be improved and all constants can be
given explicitly as expressions containing products of Gamma functions from
Selberg's integral.

\begin{figure}[t]
\begin{center}
\includegraphics[width=7cm]{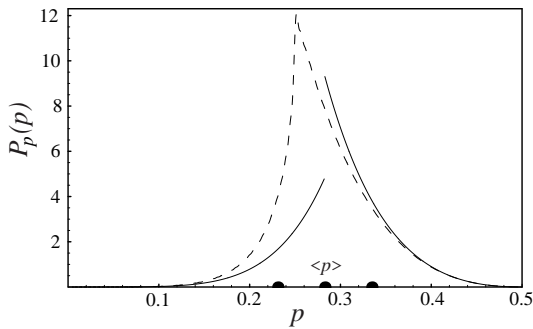}
\caption{\label{fig3} The distribution of the shot-noise power (dashed line)
and corresponding asymptotics (full lines) at $\beta = 4$ and $N_1=4$,
$N_2=2$.}
\end{center}
\end{figure}
In figure \ref{fig3}, we show as an example the distribution of the
shot-noise power at $\beta=4$, $N_1=4$, $N_2=2$ (dashed line) with
calculated asymptotics near the edges (full lines). Indicated as dots are
the average $\langle p \rangle$ and $\langle p
\rangle\pm\sqrt{\mathrm{var}(p)}$ known exactly. We have also compared (and
found a perfect agreement) all the expressions with numerical simulations of
the RMT statistics of $T_i$ and resulting distributions for $g$ and $p$.

\section{Conclusions}

In summary, we have applied essentially the theory of Selberg's integral to
the problems of quantum transport in chaotic cavities. The cumulants up to
the forth order for the conductance and up to the second order for the
shot-noise power have been calculated exactly at arbitrary channel numbers
and repulsion parameter $\beta$. We have also given the conductance and
shot-noise distributions in closed form suitable for subsequent analytic
analysis (e.g., finding asymptotics) as well as for numerical
implementations. It would be desirable to compare our findings with relevant
experimental results, e.g., in microwave cavities which became available
recently \cite{Hemmady}. This could, however, be not so straightforward, as
it requires taking into account effects of dephasing \cite{Brouwer} and
absorption \cite{Fyodorov} as well. Further work in this direction is
needed.

Support by SFB/TR12 of the DFG is acknowledged with thanks.

\bibliographystyle{unsrt}

\end{document}